# Supporting the Supporters of Unaccompanied Migrant Youth: Designing for Social-ecological Resilience

**Franziska Tachtler**
TU Wien
Vienna, Austria
franziska.tachtler@tuwien.ac.at

**Toni Michel**
TU Wien
Vienna, Austria
toni.michel@tuwien.ac.at

**Petr Slovák**
King's College London
London, UK
petr.slovak@kcl.ac.uk

**Geraldine Fitzpatrick**
TU Wien
Vienna, Austria
geraldine.fitzpatrick@tuwien.ac.at

**ABSTRACT**

Unaccompanied migrant youth, fleeing to a new country without their parents, are exposed to mental health risks. Resilience interventions mitigate such risks, but access can be hindered by systemic and personal barriers. While much work has recently addressed designing technology to promote mental health, none has focused on the needs of these populations. This paper presents the results of interviews with 18 professional/ volunteer support workers and 5 unaccompanied migrant youths, followed by three design workshops. The results point to the diverse systems that can facilitate youths' resilience development. The relationship between the youth and volunteers acting as mentors is particularly important for increasing resilience but comes with challenges. This suggests the relevance of a social-ecological model of resilience with a focus on designing technology to support the mentors in order to help them better support the youth. We conclude by mapping out the design space for mentor support.

**INTRODUCTION**

Many children and young people must flee their home countries without their parents. The term "unaccompanied minors" describes *"all foreign nationals or stateless persons below the age of 18, who either arrive in the EU unaccompanied by a responsible adult or who are left unaccompanied after their arrival"* [42, p. 7]. This highly diverse group with different ethnic backgrounds and varying motivations for fleeing their country are exposed to the risk of mental illness due to their experiences before, during, and after their flight [23, 30, 33]. These include issues such as arriving in a country of asylum, overcoming cultural differences, lacking access to supportive social relationships, handling the difficult transition to adulthood, and dealing with an uncertain future without parental support [32, 33]. When they turn 18, the difficulties become further complicated by the loss of legal protection as minors.

Thus, unaccompanied migrant youth could benefit from mental health support to guide them in developing strong resilience skills. Psychological resilience refers to the ability to adapt *"in the face of adversity, trauma, tragedy, threats or stress"* [6] and reduces negative mental health outcomes [77]. However, unaccompanied migrant youth face systemic and personal barriers to accessing such support [24, 28, 34, 44] as well as resilience-building interventions in general. While there is growing evidence that technology can support the delivery of mental health interventions (e.g., online therapy [68] or apps [3, 22, 25, 54]), there is a lack of mental health technologies that would directly address the needs of refugees.

To understand how the current support structure promotes resilience, we conducted interviews with 5 unaccompanied migrant youths as well as 18 professional and volunteer support workers spread across social work, educational programs, mental health promotion, and mentorship programs. We learned that, in contrast to professional support workers, volunteers acting as mentors build up an important trust relationship with the youth. While this could provide a good basis for delivering mental health interventions, there are challenges because they feel overburdened with their role. To deepen our knowledge about the mentors' challenges and support practices, we conducted three co-design workshops with volunteers. The results showed the ways in which the mentors wanted more support for developing mental health expertise and address network-related challenges that hamper their ability to strengthen unaccompanied migrant youth's resilience. Thus, supporting the mentors presents an area of significant potential for technology-enabled support, which in turn could lead to better resilience support for the youth.

This support direction aligns with research in psychology that examines resilience promotion for young people from a more theoretical perspective, namely the social-ecological model of resilience [63, 64, 65, 66]. This model provides conceptual grounding for the proposed shift from viewing resilience development as a responsibility of the individual, namely youth, to promoting and designing an environment which supports resilience development. In addition, this model articulates the interactions, attributes, and interplay between systems that are required to provide a support structure promoting resilience. We use this to map out the design space for how technology could promote resilience in unaccompanied migrant youth by supporting the supporters. We focus on one specific support group – the mentors – as well as the systems in which they are involved. The opportunities for design are important not only for the unaccompanied migrant youth context but also for the broader care context regarding complex care networks and needs (e.g., children with mental illnesses [39], complex conditions [4] and special needs [5]) and support work in humanitarian crisis response (e.g., [56, 58, 60]) and voluntary work (e.g., [20, 21]).

## BACKGROUND AND RELATED WORK

### Resilience and unaccompanied migrant youth

Unaccompanied migrant youth experience increased mental health risks – compared to both non-migrant youth and refugee children with parents – and display higher levels of post-traumatic stress disorder symptoms [33]. However, resilient children possess certain personal attributes that prevent the onset of mental health problems. These attributes are social competence, problem-solving skills, critical consciousness, autonomy [9, 10], and a sense of purpose [10]. Research has identified other risk and protective factors besides personal attributes. Risk factors inhibit resilience and include racial discrimination, minority status, and negative life experiences as examples [77]. Conversely, protective factors promote resilience [77] and include coping mechanism, emotion regulation, and self-management. These protective factors can be promoted through preventative interventions which encourage resilience [50]. By integrating these exercises into everyday life (e.g., teaching identifying and labeling emotions and managing anxiety [36, 40]), unaccompanied migrant youth could develop their resilience and thereby more effectively cope with their circumstances.

However, unaccompanied migrant youth face difficulty accessing services and might not trust them due to their experiences and backgrounds [24, 34, 44]. Cultural differences regarding healthcare [28, 34, 44] as well as attitudes towards mental illness [44], language [28, 34], fear of being arrested [28], and lack of knowledge of rights [28, 34] represent additional factors hindering access to health services.

### Refugee and HCI

A growing body of work in Human-Computer-Interaction (HCI) has investigated how technology can support the refugee population. Some projects indirectly support the wellbeing of refugees by decreasing risk factors and increasing protective factors. For instance, risk factors can be decreased by addressing complex issues during the post-migration phase (e.g., by providing technological aids for accessing healthcare services [52, 59], overcoming language barriers [15] or accessing services in a new country [18]). Other technological aids increase protective factors by supporting the resettlement process and integration into the host community to rebuild social capital [1, 2], and cultivate supportive community structures [69]. Few HCI projects focus specifically on supporting refugee children and young people by empowering them [26, 73] and creating a safe space for self-expression where they can be heard through participatory engagements [13, 16], which indirectly contributes to resilience. Brown et al. [13], for example, investigated how HCI contributes to mental health promotion in the refugee context, by analyzing how co-creation processes foster post-traumatic growth. However, while we can examine these projects in terms of resilience, none have specifically designed mental health technology with refugees' needs and their broader context in mind.

### Mental health technology

Increasing evidence shows that technology could increase accessibility to mental health interventions [3, 22, 25, 54] and there are many commercially available apps, e.g., Wysa [72], Woebet [71], Headspace [29], and SuperBetter [57]. It has also been shown that online services can make therapy more attractive and accessible for young people [68]. However, while many studies have focused on access difficulties, also an issue for unaccompanied migrant youth, none have examined specifically the needs of refugees and there is only limited work on how to support people caring for youth with mental illness [39, 75]. We therefore investigate how existing support structures promote resilience in unaccompanied migrant youth in order to identify opportunities for technology-enabled support. Before presenting the study's methods and findings, we first provide an overview of the study context.

## OVERVIEW OF THE STUDY AND CONTEXT

Our study took place in Austria and we focused on the post-migration phase because of its crucial impact on psychological outcomes [24]. We provide here an overview of the asylum application procedure and local regulations due to their impact on everyday life and support structures.

Unaccompanied migrant youths have to undergo an asylum application process that can take several years, during which they must attend several hearings to prove their right to asylum. After each hearing, the youth is given an asylum status which can either result in asylum, subsidiary protection, or a negative answer. In cases of a negative answer, they can take legal steps to try to prevent their deportation [7, 8].

In their everyday life, unaccompanied migrant youth interact with different types of professional support workers. Everyone who arrives in Austria without permission must remain at a collection center until officials determine jurisdiction. If a person claims to be a minor, they must undergo an age determination process. If they are determined to be aged between 14 and 17, they are assigned to a residential home for unaccompanied minors [7], where they are allowed to stay until they turn 18. These residential homes are organized by non-governmental (NGOs) and non-profit-organizations (NPOs). Depending on their age, status, and proof of education, unaccompanied migrant youth have access to different educational programs, job trainings, or the job market. As unaccompanied migrant youth flee without parental support, some NGOs and NPOs organize volunteer support workers to act as mentors through different types of mentorship programs.

| Profession | Number | Participant code |
|---|---|---|
| **Social workers** | | |
| Homes < 18 | (4) | SW2 SW3 SW5 SW6 |
| Follow-up care | (1) | SW1 |
| School | (1) | SW4 |
| **Teachers** | | |
| Project leader | (1) | T1 |
| Liaison teacher | (1) | T2 |
| Photo voice WS | (1) | T3 |
| **Mental health experts** | | |
| Psychotherapist | (1) | P1 |
| Community worker | (1) | P2 |
| Clinical psychologist | (1) | P3 |
| **Mentorship** | | |
| Coordinators | (3) | C1 C2 C3 |

Table 1. Overview of the professions

**METHODS**

We use qualitative methods to achieve a deeper understanding of the context, and to help identify potential directions for how technology could serve specific needs.

**Data collection**

*Interviews*

We interviewed 15 professional working in different areas to gain diverse perspectives on the support structure and everyday life of unaccompanied migrant youth (see professions in Table 1) and 3 volunteer support workers acting as mentors (M1-3). We recruited all interview partners in Vienna through NGOs and NPOs. Participants are referred to as SW for social worker, T for teacher, P for mental health expert, M for mentor, Y for unaccompanied migrant youth, and C for program coordinator, all followed by a participant number. Two interviews (P2, P3) occurred in London where we recruited through conferences on mental health and refugees. One teacher (T3) facilitates photography workshops called "PhotoVoice" to empower female unaccompanied migrant youth. One psychotherapist (P1) offers anti-aggression workshops for refugees. The clinical psychologists (P3) and a community worker (P2) facilitate wellbeing workshops. The program coordinators (C1-3) have organized and supervised many different mentoring relationships. One coordinator (C3) was also a mentor, and one social worker (SW5) also wrote a Master's thesis on the mentoring relationship.

We also interviewed 5 unaccompanied male migrant youths (18-21 years old) from Middle East and South Asia (3), Horn of Africa (1) and Northern Eurasia (1) (Y1-5). We only recruited young people 18 and older so that they could decide to participate in the study without an assigned legal guardian deciding for them. We aimed for gender-balance, but it was not possible since most unaccompanied migrant youth are male. During the recruitment process, NGOs and NPOs reported that female unaccompanied migrant youth are difficult to reach and rarely participate in public life. Two youths still lived at a residential home for unaccompanied minors and were in the process of moving to a shared flat for unaccompanied migrant youth above age 18. The rest lived in self-organized shared homes. Some of the young people had received asylum, some had received subsidiary protection, some had received a negative answer, and others were awaiting their hearing.

The interviews were semi-structured and featured questions broadly covering topics including everyday life of unaccompanied migrant youth, available resources, challenges, and use of technology. For all interviews, the time and place were chosen by the interviewee to make them as comfortable as possible. Two mentees were interviewed together with their mentor. Two youths living at the same residential home for unaccompanied minors wished to be interviewed together. And two program coordinators (C1, C2) were interviewed together. The rest of the interviews were one-on-one. Each interview lasted between 40 to 60 minutes and was conducted by the first author. The second author attended some interviews. Both the first and second authors took notes as well as discussed their notes immediately following each interview. Four of the five youths preferred their interview not be audio recorded, while all other interviews were audio recorded and transcribed.

To further protect the youths' anonymity, we do not present any contextual information which could make them identifiable. All necessary ethical procedures were followed as required by the host universities. No work was conducted until review by ethics advisory boards in Vienna and London had occurred.

*Workshops*

Following the interviews, we conducted three co-design workshops with mentors, framed as being about developing a guidebook for new mentors. This co-design approach provided a constructive forward-looking way both to gain a deeper understanding of the common challenges faced by mentors as well as to enable them to articulate and capture support approaches they themselves found successful. Such a participatory, explorative approach [67] also connected with their intrinsic "volunteer" motivation to help and shifted discussions from problems to potential solutions. We invited the interviewed mentors and SW5 who wrote his thesis on the mentorship relationship. We recruited additional mentors via different mentorship programs. Table 2 lists the different mentorships, also showing the mentors' rich experiences.

In the first workshop, participants (M1, M2, M4, M5, M6, SW5) brainstormed problems and advice and grouped them into categories, while in the second workshop participants (M7, M8) mapped out phases of the mentoring relationship and categories of advice onto a timeline. In addition, this workshop focused on topics which emerged during the first workshop, namely regarding expectations, setting boundaries, and discussing mental health. At the end, participants developed ideas for technology-driven gratitude interventions as an exemplar of an intervention that can help promote resilience. Based on the outcomes of the workshops, the first author drafted a guide for newcomer mentors, which was presented and discussed in a third workshop, attended by five mentors (M1, M5, M6, M7, M8) and the social worker (SW5). In groups, they edited the draft for the guide for newcomer mentors and then discussed their results with the whole group.

| Mentor | Age of the mentor-mentee relationship | Mentee's age |
|---|---|---|
| M1 | 3 years | 19 years |
| M2 | 5 years | 21 years |
|    | 2 years | 18 years |
| M3 | 10 years | above 18 |
|    | 2 years | 20 years |
| M4 | 1.5 years | 17 years |
| M5 | 3 years | 18 years |
| M6 | 3 years | above 18 |
| M7 | 11 years | 24 years |
|    | 3 years ago, lasted 1.5 years | 20 years |
|    | 1 year | 20 years |
| M8 | 4 years | above 18 |
|    | disappeared after 1.5 years | under 18 |
|    | unknown | above 18 |

Table 2. Overview of mentors' experience with their mentees

**Thematic Analysis**
We used a constructivist approach to achieve a deeper understanding of the specific context and peoples' perspective of the situation [43]. We conducted a thematic analysis [11] of the transcripts and notes from the interviews that were not recorded, which we inductively coded using software and then iteratively clustered into groups. The first author was responsible for coding. Several brainstorming meetings were held with the other authors to discuss groups of codes and potential themes, and manually clustered the codes together. During the coding process, the first author developed several thematic maps to better understand the different relationships between different groups of codes. We learned that our findings from the thematic analysis are closely aligned with research that examines resilience development from a more theoretical perspective, namely the social-ecological model of resilience [63, 64, 65, 66] which we present in the discussion section.

**FINDINGS**

**Everyday challenges and the role of supporters**
When we asked the youths about their everyday life in the interviews, they all discussed learning and studying. They also declared a desire for certain activities such as learning programming, helping people, and engaging in joint activities, but they lack time to do so. A social worker explained that one reason that education is so dominant in the youths' everyday life is that the ones *"who are top-performing have the biggest chances to stay"* (SW1). There are a limited number of open jobs and school programs and the youth can only access these if they earn sufficient grades, possess German skills, and only until they reach a certain age. Educational programs that target youth with a refugee background have *"a huge waiting list"* (SW4). The youths must handle this performance pressure under completely different circumstances from youth in the host nation: *"30 percent do not know the alphabet"* (SW2). When they turn 18, unaccompanied migrant youth *"only receive basic social services"* (SW2). Because some families expect their youth to send money home, the youth feel even greater pressure to quickly finish school to earn money.

Different organizations connect the unaccompanied migrant to a network of supporters who care for them professionally. Each professional focuses on their field of expertise, such as teachers on education including teaching literacy, German, or basic education. Social workers at residential homes or educational institutes ensure that young people become autonomous and progress in their personal development such as by enrolling them in a course, managing the asylum procedure paperwork, and preparing them for upcoming hearings. Both teachers and social workers attempt to provide basic mental health support. Teachers try to create a safe space where *"there is as little pressure as possible"* (T3) and the youth *"can forget all their problems for a short moment"* (T1), *"be themselves"* (T2), and *"make decisions"* (T3). Social workers try to *"give moral support"* (SW3) by increasing awareness of the youths' achievements, asking them how they feel, and *"showing that (they are) there for them"* (SW6). However, there are several gaps in the infrastructure of professional support workers, which we present in the next section.

**Gaps in the professional care infrastructure**
Political regulations limit the ability of professional support workers to offer a care environment which fulfills all needs of unaccompanied migrant youth. First, the care ratio is too low, as these young people *"need intensive care, not a 1:15 care"* (SW2). Due to the lack of human and financial resources as well as the significant amount of time-consuming administrative work, professionals lack time to provide sufficient support for the young people as there is insufficient time for 1:1 meetings. Second, the young people lack a secure long-term perspective of their care situation, as the youth must move out when they turn 18 due to political regulations. Consequently, the social workers avoid developing a 1:1 trust relationship with the youth in order to ease the separation. However, a close relationship forms the basis for discussing mental health and delivering mental health interventions, as SW3 remarked, *"a change only happens within the basis of a relationship"* (SW3). For example, before motivating the young people to shift from black-and-white thinking to more positive thinking, a relationship with the youth must first be built. Only then can conversations *"happen on an emotional level"* and the young people *"understand for what they change things"* (SW3). Professional support workers cannot change these circumstances and must handle poor *"circumstances which (they) do not set, but are decided for (them)"* (SW3).

Another substantial difficulty regards engaging young people earlier and more regularly with mental health interventions outside of the clinical context. According to the interviewed mental health experts, this is needed to stabilize the youth and prevent their need for professional help. The young people struggle to *"accept (psychological) support"* (C2), as mental health represents a *"stigmatized term"* (P2) for unaccompanied migrant youth. In their home countries, mental health entails a *"negative connotation"* (SW4) and people *"get committed to an insane asylum"* (SW2). Even when interventions are offered as group workshops, which seem to be more accepted, the experience of a therapist is that ultimately only *"model students"* and *"not the ones who need it the most"* (P1) attend them. Even when the youth accept psychological support, one challenge is that they do not attend regularly, which is important to strengthen resilience. For instance, a young person *"found a therapy place (and then) only went to the therapy twice before quitting it"* (SW6). Reasons are that the youth *"have a lot of appointments and then attending therapy additionally gets too much"* (SW2) and *"many believe they go once to the therapist and all their problems are gone"* (SW2). In addition, during the asylum procedure, the youth are *"not mentally resilient and it is easier to repress things"* (SW2).

**Important role of mentors**
Professional support workers attempt to connect the youth with a volunteer acting as a mentor before the youth turns 18 and must leave the sheltered home, *"as otherwise the risk is high that the youth plunges into a deep mental abyss and descends (...) into the drug scene"* (SW3). The youth who did not have a mentor shared his *"wish to have a mentor"* (Y3). Unlike professional support workers, mentors care for youth based on a trust relationship and provide individual support. Mentors meet the mentees regularly in their everyday life and remain in constant contact via phone calls or messages. Even if the mentees live an independent life, they still contact their mentors when something important arises such as extending subsidiary protection: *"If my mentee needs something, e.g., recently extending the subsidiary protection, it is very clear that we do that together with him"* (M6). Mentors offer a safe place that young people can access when they *"need extra support"* (M3), *"where they can be how they are"* (SW5), *"can catch up their lost puberty"* (M7), and *"admit that it is currently not such an easy time and that they cannot deal with certain roles"* expected by their family (C1). One interviewee compared this relationship to an *"anchor"* which prevents the youth *"from disappearing day to day and maybe disappearing completely"* (SW4). One mentee stated that the mentorship felt strange at the beginning but that they *"grew together"* (Y4). Another mentee explained that *"the time, when they did things together to be polite to each other, is over"* (Y1).

**Mentors' challenges with providing good care**
As described in the section regarding professionals' challenges, a highly functioning and trusting relationship forms an important basis for discussing mental health and delivering mental health interventions. However, mentors feel overwhelmed and overburdened by their role to provide mental health support, as they are not mental health experts. We identified several gaps and challenges, grouped into the following categories: the self, the relationship between mentors and mentees, and network-related challenges.

*Self: Dealing with their own expectations*
We identified several struggles regarding the self and wellbeing; the mentors struggle with feeling significant responsibility and high expectations towards themselves that are difficult to fulfil due to the relationship's complexity, numerous external challenges, and the youth's mental ups and downs. During and after the relationship-building phase, mentors must become experts in specific topics such as law or education in a brief time span, and they must navigate frequently changing rules and aid projects. The requirement of working across different areas of expertise makes it difficult to fulfil the high expectations mentors hold for themselves. Even if some mentorship programs provide a framework for mentors' tasks by how they describe the mentorship, such as by classifying them as a trust, learning, or work mentor, mentors must define and clearly communicate what they can offer as a mentor, where their boundaries are, *"define clear intentions"*, and ensure that they *"do not leave room for misunderstandings"* (C2) such as cultural misunderstandings.

It is difficult for mentors to establish boundaries and accept that it is not possible to solve everything, especially if the mentees do not feel well. However, it is important for mentors to set boundaries for their own mental health due to the close nature of the trust relationship, as mentors are exposed to empathy stress and can feel overstrained by mentees' severe situations and emotional ups and downs. In the workshops, the mentors described the mentorship challenges as numerous ups and downs occurring in an endless loop. Being a mentor can be *"extremely exhausting"* (M2). One mentor had to frequently explain to her mentee that she does *"not have the energy to pull him out of his deep well"* (M8). Mentors are confronted with young people's uncertain situations and must also manage the threat of deportation of someone they have grown to love in the instance of a negative answer during the asylum procedure. One mentor explained that she is *"worried about what is happening at the (asylum) hearing"* (M3), while another explained that while awaiting the decision on her mentee's asylum applications, he told her *"before he gets deported, he rather kills himself"* (M2), which was *"quite shocking"* (M2). One mentor, whose mentee experiences problems with the police, shared that she *"cannot stop talking about how much stress she currently has with"* (M1) her mentee.

*The relationship between mentors and mentees*
Another challenge for mentors regards providing direct mental health support, which can be grouped into three categories: reading mental health symptoms, discussing mental health, and providing support.

The mentors at the workshop suggested that other mentors should closely observe their mentees in order to identify changes such as *"uncertainties in decisions and the decision-making processes,"* (SW5) as well as *"mood swings, and fluctuations in decisions"* (M8). However, according to experienced mentors, understanding their mentees' mental health represented a learning process: *"It is a learning process to try

*to always reflect if the young people mention physical things, that this may be an expression of a psychological problem, as for us (adults), it is much easier to name things"* (SW5). Some young people contact their mentors when they feel unwell, such as by phoning them, and then speak to them about their wellbeing in an indirect way. For example, they may talk about physical symptoms such as general pain, stomach pain, insomnia, or nightmares. They may use words such as *"my head is broken"* (M7, M8) and *"losing temper"* (M7) or explain that they *"have stress"* (M1) or *"sleep a lot"* (M8). One mentee explained that he practices Taekwondo *"to get rid of his aggression"* (Y1) while another draws pictures as he has *"many problems from his brain and head (...) and (he has to) do something so they go away"* (Y5), as he had *"a lot of thoughts about the past"* (Y5).

Even after developing a close relationship with their mentor, the young people struggle with discussing their mental health and feel ashamed, believing they must be a strong man or role model. One mentor explained that for her mentee, *"it is not easy to not be perfect, not to be an absolute role model, like a mannequin"* (M4). Another mentor described how her *"first (mentee) partly hid himself, then he disappeared, and then he came back. But that was maybe a matter of shame – he always wanted to be a strong man"* (M8). To overcome this taboo topic, one mentor (M4) tried to create a code to discuss mental health problems, as she knows that when her mentee does not contact her, it means that he does not feel well, and thus she asks if he would like to visit to hold her newborn. She knows that if he agrees, her suspicions would be confirmed.

Currently, mentors promote mentees' wellbeing by planning common activities (e.g., going for a walk) or by ensuring that their mentees have a structured daily life and need to rise in the morning (e.g., by enrolling them in a course or planning trips during the school summer break). In addition, mentors empower their mentees to break free from their role as a victim and receiver of help by requesting their help with activities. One interviewed youth proudly explained that he *"cooked for (his) mentors"* (Y4) while another helped his mentors *"to dig up the garden"* (Y1). However, mentors lack practical advice and resources to support their mentees' mental health that are easy to apply and useful in the long term. One mentor criticized the draft guidebook for newcomer mentors, which was developed during the first two design workshops and presented at the third, for lacking concrete advice that could help in the long term: *"Which advice do I give a depressive person? Yes, to go for a walk today, but this doesn't help – better would be advice about which therapy possibility he has"* (M5). The advice by some workshop participants to know their limits and connect the young person with experts fails to succeed in practice when the mentee refuses to see a therapist, which is usually the case as mentees often instead tell mentors that *"it is enough if I talk with you"* (M2) or *"How does talking (with a therapist) help? Nobody has to talk"* (M6).

One mentor even attempted some calming exercises with her mentee, which she found online from a book for helpers about dealing with trauma, such as by drawing mandalas. However, the mentee rejected the exercises, claiming that he was *"no longer a child"* (M7). When the mentors brainstormed technological tools for their mentees to record positive thoughts daily, they emphasized that an activity must be simple to perform since their mentees are often tired, but it also must be connected to their interests such as games, music, or animals. One mentor argued that there also needs to be space for negative thoughts: *"It is important that they get rid of their worries, as otherwise you (the mentors) become their trash bin"* (M8). In addition, mentors feel isolated and desire the opportunity to easily contact experts, for example to check whether they were correctly interpreting their mentees' behavior. During a workshop, one mentor suggested that an anonymous online chat to reach out to experts would be helpful for both mentees and mentors. However, language and slow typing skills might pose barriers to mentees in using the chat.

*Network related challenges: Coordinating care*
For the mentors, providing help is easier when there is effective coordination and communication between different professionals and volunteers who care for the same youth. Otherwise there could be conflicting methods of providing support and the young person might engage in "caregiver shopping" which is counterproductive to developing a trust relationship. For example, one mentor (M6) described a scenario where she was attempting to motivate a young person to attend school while another unknown volunteer, with whom she lacked contact, tried to help the young person to find a job. At the workshop, mentors stated that at the beginning of the mentorship it was difficult to form a clear understanding of who was providing which support and how, since there were many actors from different organizations involved in the care structure. Especially in the beginning, mentors struggled to find concrete offers and help with specific needs such as legal and social assistance. Different aids are offered by NGOs that only exist with sufficient funding and are only accessible to youths who have been granted asylum. Experienced mentors suggested that the guidebook for newcomer mentors should *"give (the newcomer mentor) the idea that there is someone else"* (M7) *"and maybe show a person with a question mark because often there are people who you do not know anything about"* (M1). In addition, they suggested advising newcomers to research available programs for unaccompanied migrant youth.

Effective coordination can increase the quality of care and minimize mentors' challenges with providing support in many different areas of expertise. For example, one mentor stated she was glad that there is someone who focuses on her mentee's overall educational development and another who teaches math, as she would not be proficient at that. According to her, *"it is important that everyone works along similar lines and knows that the roles are distributed"* (M4). When sharing with professionals about their mentees' mental health, mentors shared concern that they are unsure about what they are allowed to know (e.g., due to privacy issues). On occasion, social workers communicate information about a youth – without that person's knowledge and permission – to support the mentor. For example, the social worker warned the mentor (M4) that the mentee consumed medications for mental health issues and thus may be somewhat absent even though he was fine, so that the mentor knew how to react to the behavior.

*Network related challenges: The exchange between mentors*
Unlike professional support workers, the support structure for mentors is limited. Professionals work in a team, regularly share effective strategies for distancing themselves, check on each other and their wellbeing, and hold regular meetings with their team and a supervisor. In some mentorship programs, the mentors start together in a small group and meet a few times in the beginning. The program coordinators, who are busy and struggle with funding problems, try to be present for questions and to organize opportunities for mentors to share their experiences. Some programs offer monthly meetings and meet-up events such as a yearly picnic or exhibition. However, according to the program coordinator, these meet-ups are not well attended. One reason could be that some mentors do *"not engage in the mentorship to get to know other"* mentors (M2). In addition, at the meet-ups mentors *"talk less about problems in the mentorship"* (M1), as these are complex and it requires time to *"understand the situation and questions of someone else"* (M1). Thus, there are limited exchanges regarding similar challenges such as mental health issues, upcoming hearings, and finding jobs for their mentees.

Sharing between mentors seems to be highly beneficial, especially between new and experienced mentors but also for mentors who encounter similar situations. According to one mentor, *"it is very supportive to have a 1:1 exchange with other mentors"* (M1). At the workshop, experienced mentors emphasized the value of using group reflection instead of self-reflection to define their boundaries. This study's co-design workshops also demonstrated that sharing among mentors across networks could be fruitful and positive. For example, mentors exchanged reading tips for helpers about supporting traumatized children and advice regarding relevant local organizations that were preparing and assisting their mentees with finding a job. At the end of the workshop, the mentors were thankful for the opportunity to discuss with others about the challenges and positive experiences of the mentorship.

## DISCUSSION

In this paper, we investigated how current support structures promote resilience in unaccompanied migrant youth. We learned that different groups of supporters are involved in encouraging resilience and that volunteers acting as mentors play a key role in supporting unaccompanied migrant youth due to developing a 1:1 trust relationship. However, these mentors face many complex challenges in supporting the youth.

In seeking to make sense of these findings and address the entailed challenges, we argue for the value of a social-ecological model of resilience [63, 64, 65, 66]. First, it provides a theoretical account of our findings and a framework for designing resilience promotion from an ecological rather than an individual approach (in this and similar contexts). Second, it specifies the different interactions, attributes, and interplay between systems which characterize the social-ecological model of resilience, thereby providing a framework to identify potential pathways of technological interventions to promote resilience in unaccompanied migrant youth. Third, it provides a conceptual structure to inspire novel design solutions by drawing analogies to existing work from different contexts that have targeted similar interventions or support models.

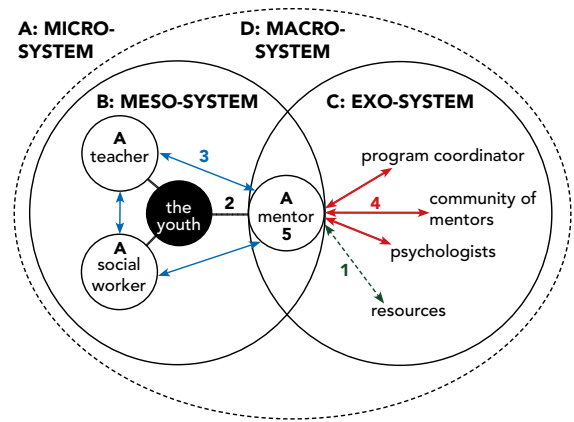

Figure 1. Social-ecological model of resilience with the focus on the mentor-mentee relationship and associated support systems. The numbers pinpoint identified challenges, termed Cases (see Section The social-ecological model of resilience and Section Design directions).

**The social-ecological model of resilience**
In a social-ecological model of resilience, the environment is most facilitative of resilience if (i) the individual, namely the supporters and the supported young person, can easily navigate and access resources, and apply appropriate and culturally meaningful resources and expertise; (ii) the exchange inside and across systems functions well; (iii) solutions as well as the supporters' wellbeing and capacity to support is supported for long-term sustainability [65, p. 388].

This definition of the social-ecological model of resilience provides a framework for deepening our understanding of the challenges the mentors face when supporting their mentees. The mentors struggle with providing mental health support as they encounter challenges across all these aspects above: (i) They struggle finding and accessing resources (Figure 1, Case 1) and expertise (Figure 1, Case 4). Even when the mentors manage to find resources, they are insufficient to meet the mentees' needs and the mentors struggle with promoting the wellbeing of their mentees in the long term (Figure 1, Case 2). Thus, the characteristic of the social-ecological model of resilience, where individuals can easily navigate resources that they can apply to their specific situation, is not fulfilled in this context. (ii) There is a clear gap between the coordination of care (Figure 1, Case 3) and the sharing between mentors caring for different youths (Figure 1, Case 4), which decreases the quality of care. Thus, the exchange between different systems does not function well. (iii) The mentors struggle with empathy stress, high expectations towards themselves, and feel overburdened by their role to provide mental health support. Their own mental health is at risk which threatens their ability to provide stable support in the long term (Figure 1, Case 5).

**The interplay of systems and pathways for interventions**
The social-ecological model of resilience provides lenses to help structure pathways of interventions and to map out application areas for technological support (see systems in Figure 1). We can interpret our findings through the lens of the different systems identified by Bronfenbrenner's ecological theory of development for children [12]: bio, micro, eco, meso, macro, chrono [66].

In the scope of this paper, we emphasize the three systems in which the mentors play an active role: In micro-systems (A), the young person, in this case the unaccompanied migrant youth, is directly involved (e.g., mentorship and residential home) [66, p. 352]. Meso-systemic processes (B) describe the interactions between micro-systems. The quality of interactions between the mirco-systems contributes to a better mental health outcome (e.g., an effective exchange between mentors, social workers, and teacher to support individuals' development) [66, p. 354]. The exo-system (C) concerns the many different distal social interactions that indirectly influence the quality of meso- and micro-systemic interactions and therefore affect young people's resilience (e.g., effective community support for mentors positively influences the quality of the mentor-mentee interaction) [66, p. 354-355].

**Supporting resilience through supporting the mentors**
Other studies in psychology further ground the argument that supporting the young persons' mentors leads to mental well-being in the young person. Supportive environments, good care quality, social support and the feeling of being cared for all lead to better mental health (e.g., [23, 30, 55]). In addition, promoting relationship duration, structure and mentor skills lead to a better quality of the mentoring relationship [51], which prevents and decreases problematic behaviours of at-risk youth [35]. Enhancing adult caregivers' wellbeing and resilience lead as well to more effective support of young people [38] (e.g., by preventing the risk of secondary trauma [70]). In addition, joint care pathways and a well-working coordination and exchange between caregivers helps to find specific solutions for the young person, which also promotes a better mental health (e.g. [23, 24]).

**Design directions**
We will now examine how technology can help promote resilience from a social-ecological model approach by looking at five specific intervention points dealing with mentors' challenges and needs (see Figure 1, Cases 1-5). Throughout our design thinking, we acknowledge that the systemic interactions across levels are complex and the boundaries between levels are diffuse and fuzzy [66]; design solutions must account for this complexity and the reciprocal relationship of dependence and influence. Interventions are needed at multiple levels. All systems stay under the influence of the macro system which includes political regulations. In the scope of this work, we however focus on the technological driven interventions and the support systems which directly include the mentors, recognising that additional support needs to be targeted at policy and infrastructure levels.

Each of the design directions supports the characteristics of a facilitative environment in the social ecological model of resilience, drawing on existing work in related areas, while emphasising the unique characteristics of this design context: the new, still developing trust-relationship of mentor and mentee.

Specifically, the mentorship relationship is positioned in-between caregiving in families and caregiving in professional settings. Nevertheless, the mentors share similar challenges as caregivers in other contexts. Many HCI projects show that informal caregiving is emotionally demanding in many different contexts (e.g., caring for children with special needs [5], family members and young people with mental illness [39, 74, 75], elderly people [53, 62] and in general [17, 41]). In contrast to caregivers caring for those with physical problems, the caregivers caring for close ones with mental illness – such as the mentors – are exposed to an even higher risk of mental illness than [39]. In addition, many HCI research projects have explored the design of systems to support carers. Examples include supporting the coordination of care (e.g., treating children with complex conditions [4], supporting young adults with autism [31]); easing the exchange of information and knowledge and organizing resources (e.g., informal caregiving of elderly [53, 62] or children with special needs [5] and voluntary work to support people with dementia [27] and who are homeless [20, 21]). Without any expertise, our mentors deal with similar challenges faced by caregivers and therapists working with young people (such as overcoming mental health stigma in the context of therapy [19] or promoting autonomy in the context of supporting young people with autism [31]). In our context, all these challenges also impact the mentors' ability to promote resilience from a social-ecological approach for the unaccompanied migrant youth. Thus, in order to develop design solutions to promote resilience by supporting the supporters, we can draw on existing solutions from other care contexts.

*Facilitate navigating resources (1)*
One important element of the social-ecological model of resilience regards mentors' capacity to navigate resources that meet their mentees' needs, are accessible, and are culturally meaningful. Several barriers hamper the mentors' ability to identify the proper resource (see Figure 1, Case 1): the resources and initiatives offered are constantly changing and scattered, many resources are available only for a specific problem and group, and the youth require specific, personalized solutions.

Technology could ease the process of navigating resources by collecting, organizing and guiding the search for relevant information. To support this process of navigating resources, the system can help in identifying which information is most relevant (such as in [20, 21] where the homeless people can ideally easily identify the latest information). In our context, mentors need to know if the resource fits to the mentee's challenge and background. A technological aid could suggest relevant and accessible information based on the mentors' and mentees' background information (e.g., location, age, and country of origin as well as asylum status) and cater the results to mentors' current challenges by asking additional questions (e.g., ask the mentor to describe the mentee's observed behavior). For instance, chatbots (e.g., [71, 72]) are designed to suggest the appropriate mental health exercises based on conversation with a person seeking mental health support.

The exo-systems in the social ecological model could support keeping the platform updated and organized. A tagging system could guide people in the exo-systems to suggest resources, assigning categories, and marking outdated resources. The program coordinators and experts could review the mentors' input. Involving and activating the community is especially important in the context of low-resource NPOs as this and in [20, 21]. Thus, the visual language has to be welcoming for non-experts to promote the sharing of content (compare [20]).

*Facilitate applying resources (2)*
In a social-ecological model of resilience, all supporters are able to apply and adapt resources to meet their specific needs and ensure that the resources are culturally meaningful. Our research indicates that the available resources, like the calming exercises tried out by one mentor with her mentee which the mentee found too childish, do not fulfil this requirement leading the mentees to reject suggested solutions (see Figure 1, Case 2). The mentors, as non-experts in mental health, desire hands-on resources that are easily applicable to their mentorship. One significant barrier to interventions is that mental health represents a taboo topic.

Technology could help mentors to apply relevant interventions. There could be a game played by both mentors and mentees that teaches the function of emotions and cultural differences in communicating about emotions. By prompting conversations about mental health the game could make mental health an integral element of daily life in the mentorship. It is also feasible for technology to help people discuss and explain mental health concepts. As an example, Coyle et al. [19] designed a computer game which eased difficult conversations between a therapist and young person by reducing stigma and making mental health concepts more accessible. However, because mentors are not mental health experts, they require additional guidance to facilitate such conversation, e.g., through online training modules (such as in [39]) which teaches how to communicate about mental illnesses.

An additional challenge for having a conversation on mental health in our context is that the mentors and mentees have different cultural backgrounds. Even mental health experts need a special training in order to work with clients from a different cultural background [37]. In addition, as the level of trust changes overtime in the mentorship and is different in every mentorship, the level of privacy of the conversation between mentor and mentee differs as well. Thus, the system supporting the mentors needs to offer different activities depending on the cultural background of the mentee and the level of trust in the mentorship. Through the exo-system, namely the network of mentors, mentors could share their experiences regarding how they applied the interventions in their mentorship.

*Coordinate care of the same young person (3)*
In our context, there is a vast number of actors, but the network is not working particularly well. Analysing this challenge from a social-ecological perspective, we identify that the meso-system, namely the system where the different micro-systems interact to provide support to the youth, does not work well (see Figure 1, Case 3). Technology which facilitates care coordination could strengthen the meso-system. For instance, a communication tool between different support workers who are responsible for one young person could potentially ease the exchange between them, make members of the care network visible, and prevent the support workers from working against each other. The system could enable creating a plan with shared care goals and ease the communication between the different actors (such as in [4]). Technology could also ease the transition phase when the youth turn 18 and usually fall-out of the care system by keeping the social worker in the technological system with a less active role.

In addition, the system could help promote the autonomy and agency of unaccompanied migrant youth by enabling them to control their own network of supporters. In [31], a system was designed to empower young people with autism to coordinate their caregivers and pose questions to different support groups. Caregivers invited trusted individuals to the network and the young person could decide on each individual's level of access [31]. One could envision a similar system that not only assigns different levels of trust to the support workers but also different types of expertise, which would empower the young person to achieve some control over their network of supporters.

*Strengthen the exchange with peers and experts (4)*
If the mentors exchange with other mentors and experts, they can provide better support and find solutions better fitting their mentees' needs. However, the existing infrastructure for this exchange does not work well due to minimal participation, the complexity of caregiving and lack of infrastructure to contact experts (see Figure 1, Case 4). Thus, the exchange between mentors and experts in this context requires support to promote resilience from a social-ecological perspective.

Technology could help connect mentors with other mentors and experts by matching based on mentees' backgrounds and struggles and facilitating both online or offline exchange. The system could recognize which expert could solve a certain problem and bring them into contact to provide more individualized support. For example, if mental health support is needed, the system could establish contact between the mentor and a mental health expert. In cases of cultural or linguistic misunderstandings, the system could locate a person with a similar cultural background. In a related HCI project [15], bringing a translator to the online communication between refugee families and volunteers helped overcome linguistic and (in some cases) cultural misunderstandings.

The system could also support the organization of local meetings based on challenges shared by many mentors. For instance, mentors who need help with the asylum procedure could suggest a meeting focused on that topic. Matching mentors by mentee's backgrounds and challenges independently of the mentorship duration would help to support the exchange between newcomer and experienced mentors and could promote a long-term commitment, which is a key challenge in online health communities [76]. In our context, the system has to be designed for different levels of technological experience. While caregivers in the age of parents might already use existing platforms (e.g., parents of children with special needs use social media [5]), older adults might have lower technology experience and benefit from systems that bridge between online and offline social support [62].

*Sustain the individual's capacity and wellbeing (5)*
An important aspect of having a resilient support network is ensuring that each individual's wellbeing and capacity to provide support is sustainable in the long term. As described in the findings, the mentors' role can be emotionally demanding due to empathy stress and the immense responsibilities they have (see Figure 1, Case 5). Coordinating the exchange between mentors and experts may help to increase the feeling of competence and consequently, the individual's capacity to provide support in the social-ecological model of resilience. In addition, research has demonstrated that simply knowing that there is an option to talk with a supervisor increases the self-efficacy of supporters [49], which may increase the wellbeing of mentors and their capability of providing support over a long period of time.

As part of the peer support system, technological features could be added to protect the mental health of the individual actor. One feature could be a request system to distribute the work load to differently experienced mentors to utilize this support structure and prevent supporters from becoming overburdened. For instance, mentors could register as having particular areas of expertise, e.g., "asylum procedure" and "job market for asylum-seekers". The request by a mentor who searches for help in a specific area could then be sent to the mentors with the relevant expertise. In addition, technology could provide venting space as well as an option to hide negative content which might effect the mentor's mental health negatively [39]; it could encourage a reflective practice through private blog posts [17] and point to other mental health promotion strategies the mentor can self-apply, and guide mentors to deal with tensions between impression management and being open about their own challenges [46].

**Broader implications**
Even if some HCI projects can be interpreted as including elements of the social-ecological model of resilience (e.g., by including social ecology in long-term mental health management [45] or by supporting caregivers coping with young people's mental illnesses [39, 75]), the social-ecological model can further be used to deepen the understanding of the different systems and inter-relationships which take a key role in providing support. The lessons from mental health promotion in unaccompanied migrant youth can also inform a broader agenda in refugee and humanitarian response and in mental health promotion and caregiving – all are areas where different systems interact to support individuals, where navigating and applying resources must be simple, and where the mental health of supporters is at risk (e.g., as in [56, 60, 70]).

Other research has also proposed using the capacity of networks via peer support systems in both the context of informal caregiving [5, 61, 62, 76] and mental health support [47, 48]. A social-ecological model of resilience further suggests that not only the systems need to be resilient but also the individual actors in the systems to be sustainable in the long term. However, peer-support systems can potentially overburden the individual. They are based on the implicit premise that people are always emotionally and educationally capable of offering support to others [14], and many show that informal caregiving is emotionally demanding (e.g., [17, 39, 53, 62]). As in case 5, technological features need to protect the mental health of the individual actor in the system, e.g., from too much negative venting. Overall, the social-ecological model of resilience provides a framework to map out the design space of technological solutions which can help the supporters (i) to deal with their challenges; (ii) to create together with the support of the different systems an environment where the young person can thrive; (iii) to stay mentally healthy in the long term.

**Limitations and future work**
In this study, we primarily focused on the potential of technology to support mentors, but this is only one direction in a social-ecological model of resilience. For instance, future research could consider the temporal dimension (chrono system) when designing technology. We could also map out the technology design space to address the challenges faced by professionals in providing support as well as the different barriers unaccompanied migrant youth encounter when accessing mental health services. As next steps, we will further develop technology concepts that are used by the mentor and mentee and that integrate resilience-building interventions in the mentorship to better understand how to design for the dynamics of this unique relationship. We also plan to speak to more youth about how mentors might better support them and to receive feedback on design ideas which more directly involve them. Overall, technology can only support these systems to some extent and there also has to be changes in policies and regulations.

**CONCLUSION**
This paper presents a first step towards understanding the role of technology in promoting resilience of unaccompanied migrant youth by supporting their mentors. We investigated how current support structures promote resilience in unaccompanied migrant youth, highlighting the important role of the mentors, their practices, challenges and needs. Our findings align with research that examines resilience development from the perspective of a social-ecological model. We used this model to draw out a range of design opportunities for how technology could potentially support mentors in promoting resilience in unaccompanied migrant youth. Although the design implications are specific to this context, the conceptual framework of the social-ecological resilience model could also facilitate mapping out the design space in other contexts with complex care systems as well as children and youth mental health areas.

**ACKNOWLEDGMENTS**
This project has received funding from the European Union's Horizon 2020 research and innovation program under the Marie Skłodowska-Curie grant agreement No. 722561.